\documentstyle[11pt,aaspp4]{article}

\begin{document}
\title{Variability of GRB Afterglows Due to Interstellar Turbulence}
\author{Xiaohu Wang and Abraham Loeb}
\medskip
\affil{Harvard-Smithsonian Center for Astrophysics, 60 Garden Street,
Cambridge, MA 02138}
%\altaffiltext{2}{email:aloeb@cfa.harvard.edu}

\begin{abstract}

Gamma-Ray Burst (GRB) afterglows are commonly interpreted as synchrotron
emission from a relativistic blast wave produced by a point explosion in an
ambient medium, plausibly the interstellar medium of galaxies.  We
calculate the amplitude of flux fluctuations in the lightcurve of
afterglows due to inhomogeneities in the surrounding medium. Such
inhomogeneities are an inevitable consequence of interstellar turbulence,
but could also be generated by variability and anisotropy in a precursor
wind from the GRB progenitor.  Detection of their properties could provide
important clues about the environments of GRB sources.  We apply our
calculations to GRB990510, where an {\it rms} scatter of $\sim 2\%$ was
observed for the optical flux fluctuations on the 0.1--2 hour timescale
during the first day of the afterglow, consistent with it being entirely
due to photometric noise (Stanek et al. 1999).  The resulting upper limits
on the density fluctuations on scales of $\sim 20$--$200$ AU around the
source of GRB990510, are lower than the inferred fluctuation amplitude on
similar scales in the Galactic interstellar medium.
Hourly monitoring of future optical afterglows might therefore reveal
fractional flux fluctuations at the level of a few percent.

\end{abstract} 


%\centerline{Submitted to {\it The Astrophysical Journal}, 1999}

\section{INTRODUCTION}

Almost all well-localized Gamma-Ray Burst (GRB) sources have shown
afterglow emission in X-rays for several hours (see Piran 1999 for a
review).  Often the emission persists on longer timescales at lower photon
energies, peaking in the optical on a timescale of days and in the radio on
a timescale of weeks or longer. This long-lasting afterglow emission is
most naturally explained as synchrotron emission from a relativistic blast
wave, produced by the GRB explosion in an external medium (see e.g.,
Paczy\'{n}ski \& Rhoads 1993, Katz 1994, 
M\'{e}sz\'{a}ros \& Rees 1993, 1997, Waxman 1997a,b). The emission frequency
declines with time due to the deceleration of the shock wave (Blandford \&
McKee 1976) and the corresponding reduction in the characteristic electron
energy and magnetic field amplitude behind the shock.  The external medium
could be either the interstellar medium (ISM) of the host galaxy (Waxman
1997a) or a precurser wind from the GRB progenitor (Chevalier \& Li 1999).

Previous theoretical models of the afterglow emission have assumed for
simplicity that the density profile of the external medium is smooth,
i.e. uniform in the case of the ISM or power-law with radius in the case of
a progenitor wind. However, this simplifying assumption is not expected to
hold in realistic situations.  The ISM is known to exhibit inhomogeneities
due to turbulence, and stellar winds may vary in time and in solid
angle. Since the emitted afterglow flux depends on the instanteneous number
of shocked electrons, any density inhomogenties are expected to induce
temporal fluctuations in the afterglow lightcurve.  Observations of these
fluctuations could provide additional constraints on the nature of the
surrounding medium and the GRB progenitor.

In this paper we derive the relation between the spatial power-spectrum of
density fluctuations in the ambient medium surrounding the GRB source and
the Fourier transform of temporal fluctuations in the afterglow flux.  We
focus on small-amplitude (linear) inhomogeneities, as those provide the
minimal source of afterglow flux fluctuations. \S 2 presents the formalism
used in our derivation, and \S 3 describes our numerical results.  We
summarize our main conclusions in \S4.  For simplicity, we consider the
case where the unperturbed ambient medium is uniform, as for a background
ISM.  The particle density inferred from detailed modeling of some GRB
afterglows is in the range $\sim 0.1$--$1~{\rm cm^{-3}}$ (e.g. Wijers \&
Galama 1998; Waxman 1997a,b), as expected for the ISM of their host
galaxies.

\section{FORMALISM} 

The relativistic fireball produced by a GRB explosion starts to decelerate
at the radius where the energy of the ambient gas swept by the fireball is
comparable to its energy output (e.g. Waxman 1997a,b),
\begin{equation}
r_c= 2 \times 10^{16}
\left( \frac{E_{52}t_{10}}{n_{1}} \right)^{1/4}~{\rm cm},
\label{eq:Waxman1} 
\end{equation}
where $E_{52}$ is the hydrodynamic energy release in units of $10^{52}~{\rm
ergs}$, $t_{10}$ is the duration of the GRB in units of 10~s, and $n_1$
is the proton density of the ambient medium in units of $1~{\rm
cm^{-3}}$. At larger radii, the shock follows the adiabatic self-similar
evolution (Blandford \& McKee 1976) whereby its Lorentz factor declines
with radius as
\begin{equation}
\Gamma=\Gamma_c\left({r\over r_c}\right)^{-3/2},
\label{eq:Waxman2}
\end{equation}
and
\begin{equation}
\Gamma_c = 270~\left( \frac{E_{52}}{t_{10}^{3}n_{1}}\right)^{1/8}.
\label{eq:Waxman3}
\end{equation}

The global spectral characteristics of GRB afterglows are naturally
explained in terms of synchrotron emission by shock-accelerated electrons
from this decelerating relativistic shock (see e.g., Wijers, Rees 
\& M\'{e}sz\'{a}ros 1997; Waxman 1997a,b; Sari, Piran \& Narayan 1998).

Figure~1 illustrates the geometry of the emission from an infinitesimal
volume element in spherical coordinates $dV=r^{2} dr d\mu d\phi$ at a
distance $D$ from the observer. Here $\mu=\cos \theta$ and $\theta=0$ along
the line-of-sight. We define the emission coefficient $j^{'}_{\nu ^{'}}$ to
be the power emitted per unit frequency per unit volume per steradian in
the rest frame of the outflowing material.  We use prime to denote
quantities in the local rest frame of the emitting material, while unprimed
quantities are measured in the rest frame of the ISM.  Note that $j_{\nu} /
\nu^{2}$ is Lorentz invariant (Rybicki \& Lightman 1979).  For a
spherically-symmetric expansion of material which emits isotropically in
its local rest frame, we have $\nu ^{'}=\nu \gamma (1-\beta \mu)$ and
$j^{'}_{\nu ^{'}}=P^{'}({\nu ^{'}},r,t)/4\pi$, where $\gamma$ and $\beta c$
are the Lorentz factor and the velocity of the emitting matter,
respectively. A photon emitted at time $t$ and place $\vec{r}$ in the ISM
frame will reach the detector at a time $T$ given by
\begin{equation}
T_z=\frac{T}{1+z}=t-\frac{r \mu}{c},
\label{eq:time}
\end{equation}
where $z$ is the cosmological redshift of the GRB and 
$T$ is chosen such that a photon emitted at the origin at $t=0$ will
arrive at the detector at $T=0$. Thus we have (Granot, Piran \& Sari 1999)
\begin{equation}
F(\nu, T)=\frac{1+z}{4\pi d_{L}^{2}}\int_{0}^{2\pi} d\phi \int_{-1}^{1} d\mu
\int_{0}^{\infty} r^{2}dr \frac{P^{'}(\nu \gamma (1-\beta \mu), r,
T_{z}+{r\mu}/{c})}{\gamma ^{2} (1-\beta \mu)^{2}},
\label{eq:flux}
\end{equation}
where $d_{L}$ is the luminosity distance to the GRB, and $\gamma, \beta,
\mu$ are evaluated at the time $t$ implied by equation~(\ref{eq:time}).

Most of the shocked material is concentrated in a thin shell behind the
shock front. The characteristic thickness of the shell is $\Delta\sim R/10
\gamma^{2}$ in the ISM frame, 
where $\gamma=\Gamma/\sqrt{2}$ is 
the Lorentz factor of the material just behind the shock.
In the following, we will assume that the
observed radiation originates from the thin shell of thickness $\Delta =
\eta R/\gamma^{2}$ behind the shock, inside of which the Lorentz factor,
the particle density, and the energy density of shocked ISM obtain
the following values,
\begin{equation}
n^{'}=4\gamma n, \ \ \ \ e^{'}=4\gamma^{2} n m_{p} c^{2} ,
\label{eq:Blandford}
\end{equation}
where $n$ is the number density of the unshocked ambient ISM in its local
rest frame, and $m_{p}$ is the proton mass.  The actual value of $\eta$
depends not only on the hydrodynamics but also on the behaviour of the
magnetic field and the shock-accelerated electrons which determine the
local emissivity.

The volume integration expressed in equation~(\ref{eq:flux}) should be taken
over the region occupied by the emitting shell at a given observed time, as
illustrated in Figure~2. Because of relativistic beaming, the
observed radiation originates from a small angle along the line of sight,
$\theta < 1/\gamma$. Hence, we can set the upper limit in the integration
over $\theta$ to be $2/\gamma_{b}$, where $\gamma_{b}$ is the Lorentz
factor of the shell at point b on the plot. For a given observed time $T$,
the outer boundary $abc$ of the integration region is defined by the
relation (Granot et al. 1999)
\begin{equation}
R_{\rm outer}=\frac{cT_{z}}{1-\mu +1/(16\gamma^{2})},
\label{eq:outer}
\end{equation}
where $\gamma=\gamma(R_{\rm outer})$.  Photons originating at this boundary are
emitted from the front of the shell and arrive at the detector at the same
time $T$ (although they are emitted at different times). Similarly, the
inner boundary of the integration region is described by
\begin{equation}
R_{\rm inner}=\frac{cT_{z}}{1-\mu +(\eta+\frac{1}{16})/\gamma^{2}} ,
\label{eq:inner}
\end{equation}
with $\gamma=\gamma(R_{\rm inner})$, and where the associated photons are
emitted from the back of the shell. Note that although the emitting shell
is thin, the region contributing to the observed flux at a given time is
rather extended; {\it this has important consequences with respect to the
spatial scale over which density fluctuations could affect the observed
flux}.  Furthermore, consider a point $g=(r,\theta)$ inside the integration
region. A photon emitted from this point at time $t=T_{z}+{r\mu}/{c}$ will
arrive at the detector at time $T$. Since the emitting shell is very thin,
the point $g$ is very close to the shock front at time $t$. Thus the radius
of the shock at time $t$ is approximately equal to $r$. This fact will be
used in our calculation.

Next, we derive the local emissivity due to synchrotron radiation.
We assume that the energy densities of the shock-accelerated electrons and
the magnetic field are fixed fractions of the internal energy density
behind the shock front, $e^{'}_{e}=\epsilon_{e}e^{'}\ ,\
e^{'}_{B}=\epsilon_{B}e^{'}$, and that the shock produces a power law
distribution of accelerated electrons with a number density per Lorentz
factor of $N(\gamma_{e})=K\gamma_{e}^{-p}$ for $\gamma_{e}\geq
\gamma_{\rm min}$, where
\begin{equation}
\gamma_{\rm min}=\left( \frac{p-2}{p-1} \right) 
\frac{\epsilon_{e}e^{'}}{n^{'}m_{e}c^{2}} , \ \ \ \ K=(p-1)n^{'}\gamma_{\rm min}^{p-1}.
\label{eq:K}
\end{equation} 
Then the emissivity can be approximated as
\begin{equation}
P^{'}=H_{1}\gamma\beta^{2} n^{4/3}(\nu^{'})^{1/3}, \ \ \ \ 
\nu^{'}<\nu^{'}_{\rm min},
\label{eq:power_2_left}
\end{equation}
\begin{equation}
P^{'}=H_{2}\gamma^{(3p+1)/2}\beta^{2}
n^{(p+5)/4}(\nu^{'})^{-(p-1)/2}, \ \ \ \ 
\nu^{'}>\nu^{'}_{\rm min},
\label{eq:power_2_right}
\end{equation}
where $\nu^{'}_{\rm min}=\nu^{'}_{syn}(\gamma_{\rm min})$ is the synchrotron
frequency of an electron with the minimal Lorentz factor, and
\begin{equation}
H_{1}=32.4\times
\frac{(p-1)^{5/3}}{(3p-1)(p-2)^{2/3}}\sigma_{T} 
\frac{\epsilon_{B}^{1/3}}{\epsilon_{e}^{2/3}}
\frac{m_{e}^{2} c^{3}}{m_{p}^{1/3}q_{e}^{4/3}} ,
\label{eq:H1}
\end{equation}
\begin{equation}
H_{2}=39.9\times1.9^{(p-1)/2}\times
\frac{(p-2)^{p-1}}{(3p-1)(p-1)^{p-2}}\sigma_{T} 
\epsilon_{B}^{(p+1)/4} \epsilon_{e}^{p-1}
\frac{m_{p}^{(5p-3)/4}c^{3}}{m_{e}^{(3p-5)/2}q_{e}^{(3-p)/2}}
\label{eq:H2}
\end{equation}
are constants. Also note that the peak frequency at which the observed flux
peaks is $\nu_{\rm peak}\sim\nu_{\rm min}$.

Subtituting equations~(\ref{eq:power_2_left}) and (\ref{eq:power_2_right})
into equation~(\ref{eq:flux}) respectively, and
making use of the fact that for a highly relativistic system, $\theta \leq
2/\gamma_{b} \ll 1$, and so
\begin{equation}
d\mu \approx -\theta d\theta, \ \ \ \ 
1-\beta\mu \approx \frac{1}{2\gamma^{2}}+\frac{\theta^{2}}{2},
\label{eq:small}
\end{equation}
we get the afterglow flux at a frequency $\nu$,
\begin{equation}
F(\nu, T)=\frac{(1+z)H_{1}\nu^{1/3}}{4\pi d_{L}^{2}} \int_{0}^{2\pi} d\phi
\int_{0}^{2/\gamma_{b}}d\theta \int_{R_{\rm inner}}^{R_{\rm outer}} dr
\frac{n^{4/3} \theta r^{2}}
{\gamma^{2/3}\left( \frac{1}{2\gamma^{2}}+ \frac{\theta^{2}}{2}\right)^{5/3}}, 
\ \ \nu<\nu_{\rm peak}, 
\label{eq:flux_1_left} 
\end{equation}
\begin{equation}
F(\nu, T)=\frac{(1+z)H_{2}\nu^{-(p-1)/2}}{4\pi d_{L}^{2}} \int_{0}^{2\pi} d\phi
\int_{0}^{2/\gamma_{b}} d\theta \int_{R_{\rm inner}}^{R_{\rm outer}}dr
\frac{n^{(p+5)/4}\gamma^{p-1} \theta r^{2}}
{\left( \frac{1}{2\gamma^{2}}+ \frac{\theta^{2}}{2}\right)^{(p+3)/2}}, 
\ \ \nu>\nu_{\rm peak}.
\label{eq:flux_1_right} 
\end{equation}
All emission frequencies under consideration are assumed to be below the
cooling frequency, $\nu_c$ (i.e. not affected by the rapid cooling of the
high-energy tail of the electron distribution), and also well above the
synchrotron self-absorption frequency, $\nu_{\rm a}$ (see Sari et al. 1998,
for more details).

At point b in Figure~2, the radius of the shock front is $R_{b}$,
the Lorentz factor of the shell behind the shock is $\gamma_{b}$,
and the time $T$ when photons emitted at point $b$ reach the observer 
is related to these quantities by
\begin{equation}
\frac{T}{(1+z)}=\frac{R_{b}}{16\gamma_{b}^{2}c} .
\label{eq:pointb}
\end{equation}
Note that this time is different from the observation time, $T_{\rm obs}$,
which is defined to be the arrival time of most photons emitted from the
shell of radius $R_{b}$. This is because most photons are emitted from a
cone of opening angle $\sim 1/\gamma_{b}$ around the line of sight and they
suffer a longer time delay $\sim R_{b}/2\gamma_{b}^{2}c$ than the photons
emitted on the line of sight (Waxman 1997c). The observation time is
\begin{equation}
\frac{T_{\rm obs}}{(1+z)}\approx \frac{R_{b}}{2\gamma_{b}^{2}c} =8
\frac{T}{(1+z)}.
\label{eq:obs_time}
\end{equation}
We can use $R_{b}$ to normalize equations~(\ref{eq:flux_1_left}) and 
(\ref{eq:flux_1_right}). 
Based on the scaling, $\gamma \propto r^{-3/2}$, we get
\begin{equation}
\frac{\gamma}{\gamma_{b}}=\left( \frac{r}{R_{b}} \right) ^{-3/2}.
\label{eq:norm}
\end{equation}
Using equations~(\ref{eq:pointb}) and (\ref{eq:norm}), we can rewrite 
equations~(\ref{eq:outer}) and ~(\ref{eq:inner}) as follows,
\begin{equation}
X_{\rm outer}^4 + 8\gamma_{b}^{2}\theta^{2}X_{\rm outer}-1=0 ,
\label{eq:outer_2}
\end{equation}
\begin{equation}
(16\eta+1)X_{\rm inner}^4 + 8\gamma_{b}^{2}\theta^{2}X_{\rm inner}-1=0 ,
\label{eq:inner_2}
\end{equation}
where $x=r/R_{b}$, $X_{\rm outer}\equiv R_{\rm outer}/R_{b}$, and $X_{\rm
inner}\equiv R_{\rm inner}/R_{b}$. In these notations,
equations~(\ref{eq:flux_1_left}) and (\ref{eq:flux_1_right}) obtains the form
\begin{equation}
F(\nu, T)=\frac{(1+z)H_{1}\nu^{1/3}R_{b}^{3}}
{4\pi d_{L}^{2}\gamma_{b}^{2/3}}
\int_{0}^{2\pi} d\phi \int_{0}^{2/\gamma_{b}}d\theta
\int_{X_{\rm inner}(\theta, \gamma_{b})}^{X_{\rm outer}(\theta, \gamma_{b})}
\frac{n^{4/3}\theta x^{3}dx}{\left( \frac{x^{3}}{2\gamma_{b}^{2}}+\frac{\theta^{2}}{2} 
\right)^{5/3}} , \ \ \nu<\nu_{\rm peak},
\label{eq:flux_2_left}
\end{equation}
\begin{equation}
F(\nu, T)=\frac{(1+z)H_{2}\nu^{-(p-1)/2}R_{b}^{3}\gamma_{b}^{p-1}}
{4\pi d_{L}^{2}}
\int_{0}^{2\pi} d\phi \int_{0}^{2/\gamma_{b}}d\theta
\int_{X_{\rm inner}(\theta, \gamma_{b})}^{X_{\rm outer}(\theta, \gamma_{b})}
\frac{n^{(p+5)/4}\theta dx}{x^{(3p-7)/2}
\left(\frac{x^{3}}{2\gamma_{b}^{2}}+\frac{\theta^{2}}{2} 
\right)^{(p+3)/2}} , \ \ \nu>\nu_{\rm peak},
\label{eq:flux_2_right}
\end{equation}
where $X_{\rm inner}(\theta, \gamma_{b})$ and $X_{\rm outer}(\theta,
\gamma_{b})$ can be obtained by solving equations~(\ref{eq:outer_2}) and
~(\ref{eq:inner_2}), and $R_{b}$ and $\gamma_{b}$ are functions of the
observed time $T$.  At a later time $\tilde{T}=T+\tau$, the shock front
moves from $b$ to $\tilde{b}$, while the Lorentz factor and radius of the
shell change to $\tilde{\gamma}_{b}$ and $\tilde{R}_{b}$, respectively,
with
\begin{equation}
\frac{R_{b}}{\tilde{R}_{b}}=\left( \frac{T}{\tilde{T}} \right) ^{1/4} ,\ \
\ \ \frac{\gamma_{b}}{\tilde{\gamma}_{b}}=\left( \frac{T}{\tilde{T}}
\right) ^{-3/8}.
\label{eq:later}
\end{equation}

Equations~(\ref{eq:flux_2_left}) and (\ref{eq:flux_2_right}) 
can be written in the generalized form
\begin{equation}
F(T)=\int G(\vec{r}) n^{y}(\vec{r})d\vec{r} ,
\label{eq:F}
\end{equation}
where the density $n(\vec{r})$ may fluctuate,
and 
\begin{equation}
y= \left\{ \begin{array}{ll} 4/3  & {\rm for}~~\nu<\nu_{\rm peak}\\
(p+5)/4 & {\rm for}~~\nu>\nu_{\rm peak}.
\end{array}\right.
\label{eq:yy}
\end{equation}
Thus we get,
\begin{eqnarray}
\langle F(T)F(T+\tau) \rangle  & = & \langle \int G(\vec{r})
n^{y}(\vec{r})d\vec{r} \int G(\tilde{\vec{r}})
n^{y}(\tilde{\vec{r}})d\tilde{\vec{r}} \rangle \nonumber \\
& = & \int d\vec{r} \int d\tilde{\vec{r}} G(\vec{r}) G(\tilde{\vec{r}})
\langle n^{y}(\vec{r}) n^{y}(\tilde{\vec{r}}) \rangle .
\label{eq:fluxauto}
\end{eqnarray}
The angular brackets in the above equation reflect an average over an
ensemble of afterglows with identical source properties, exploding at
different places in the ISM.  The ergodic assumption implies that an
average over many such systems would be equivalent to an average over 
time for the explosion if the ISM is in a stationary statistical state
(Reif 1965). For the ensemble average of the right-hand-side, we have
\begin{eqnarray}
\langle n^{y}(\vec{r}) n^{y}(\tilde{\vec{r}}) \rangle &
= & \langle n \rangle^{2y} \langle [1+\delta (\vec{r})]
^{y} [1+\delta (\tilde{\vec{r}})]^{y} \rangle \nonumber
\\ & \approx & \langle n \rangle^{2y} \langle 
1+\frac{y(y-1)}{2}\delta^2(\vec{r}) + \frac{y(y-1)}{2}\delta^2(\tilde{\vec{r}})
+y^2\delta (\vec{r}) \delta (\tilde{\vec{r}}) \rangle \nonumber
\\ & = & \langle n \rangle^{2y}
\left[1+y(y-1)\xi_{0}+y^2\xi(\vec{r}-\tilde{\vec{r}})\right] ,
\label{eq:denauto}
\end{eqnarray}
where
\begin{equation}
\delta (\vec{r}) = \frac{n(\vec{r})-\langle n \rangle}
{\langle n \rangle} ,
\label{eq:dr}
\end{equation}  
$\xi(\vec{r}-\tilde{\vec{r}})\equiv \langle\delta (\vec{r})
\delta (\tilde{\vec{r}})\rangle$ is the ensemble--averaged autocorrelation
function of the density fluctuations, and 
$\xi_{0}=\langle\delta^2(\vec{r})\rangle=\langle\delta^2(\tilde{\vec{r}})\rangle$. 
As mentioned before, we consider
only small (linear) density fluctuations. In equation~(\ref{eq:denauto}),
we have implicitely assumed $ \delta (\vec{r}) \ll 1$.

Similarly we have 
\begin{equation}
\langle F(T) \rangle \langle F(T+\tau) \rangle
=\int d\vec{r} \int d\tilde{\vec{r}} G(\vec{r}) G(\tilde{\vec{r}})
\langle n^{y}(\vec{r}) \rangle \langle n^{y}(\tilde{\vec{r}}) \rangle ,
\label{eq:fluxaver}
\end{equation}
\begin{equation}
\langle n^{y}(\vec{r}) \rangle \langle n^{y}(\tilde{\vec{r}}) \rangle
\approx \langle n \rangle^{2y}\left[1+y(y-1)\xi_{0}\right]
\label{eq:denaver}
\end{equation}

The statistical properties of the ambient gas inhomogeneities in the
vicinity of GRB sources are highly uncertain, and so we adopt the minimal
number of free parameters to describe the autocorrelation function, namely
we write
\begin{equation}
\xi(r)=\xi_{0}\exp (-\frac{r}{r_{0}}), 
\label{eq:auto}
\end{equation}
where $r_{0}$ is the scale length of the density autocorrelation function.
For simplicity, we ignore deviations of the expanding shell from spherical
symmetry. Our calculation focuses on scales much smaller than the size of
the emission region (see \S 3) and so the cummulative effect of many
small-scale patches of density perturbations averages out during the
expansion history of the shell.  At different points inside the shell, the
particle density and energy density [described by
equation~(\ref{eq:Blandford})] might be temporarily above or below their
average values, but the total energy is conserved.

We can now define the autocorrelation function of the temporal fluctuations
in the afterglow flux as
\begin{equation}
\zeta(\tau) = \langle \Delta(T) \Delta(T+\tau) \rangle =\frac{\langle
F(T)F(T+\tau) \rangle}{\langle F(T) \rangle \langle F(T+\tau) \rangle}-1 ,
\label{eq:autoflux}
\end{equation}
where we consider fluctuations on timescales much shorter than the
evolution time of the afterglow lightcurve, $\tau\ll T$, and where
\begin{equation}
\Delta(T) = \frac{F(T)-\langle F(T) \rangle}
{\langle F(T) \rangle} .
\label{eq:df}
\end{equation}

We now define
\begin{equation}
g(x, \theta, \gamma_{b}) = \left\{ \begin{array}{ll} 
{\theta x^{3}}{\left( \frac{x^{3}}{2\gamma_{b}^{2}}+\frac{\theta^{2}}{2} 
\right)^{-5/3}} & {\rm for}~~~\nu<\nu_{\rm peak}\\
 {\theta}{x^{-(3p-7)/2}
\left(\frac{x^{3}}{2\gamma_{b}^{2}}+\frac{\theta^{2}}{2} 
\right)^{-(p+3)/2}} & {\rm for}~~~\nu>\nu_{\rm peak}.
\end{array}
\right.
\label{eq:g_right}
\label{eq:g_left}
\end{equation}
One of the integrals of interest is
\begin{eqnarray}
I_{1} & = & 
\int_{0}^{2\pi} d\phi \int_{0}^{2/\gamma_{b}}d\theta
\int_{X_{\rm inner}(\theta, \gamma_{b})}^{X_{\rm outer}(\theta, \gamma_{b})}
g(x, \theta, \gamma_{b})dx \nonumber \\
 & & \times \int_{0}^{2\pi} d\tilde{\phi}
\int_{0}^{2/\tilde{\gamma}_{b}} d\tilde{\theta}
\int_{\tilde{X}_{\rm inner}(\tilde{\theta}, \tilde{\gamma}_{b})}
^{\tilde{X}_{\rm outer}(\tilde{\theta}, \tilde{\gamma}_{b})} 
g(\tilde{x}, \tilde{\theta}, \tilde{\gamma}_{b})d\tilde{x} ,
\label{eq:integ1}   
\end{eqnarray}
where $\tilde{X}_{\rm inner}(\tilde{\theta}, \tilde{\gamma}_{b})$ and
$\tilde{X}_{\rm outer}(\tilde{\theta}, \tilde{\gamma}_{b})$ are found by
solving equations~(\ref{eq:outer_2}) and (\ref{eq:inner_2}).  A second
relevant integral is
\begin{eqnarray}
I_{2} & = & 
\int_{0}^{2\pi} d\phi \int_{0}^{2/\gamma_{b}}d\theta
\int_{X_{\rm inner}(\theta, \gamma_{b})}^{X_{\rm outer}(\theta, \gamma_{b})}
g(x, \theta, \gamma_{b})dx \nonumber \\
 & & \times \int_{0}^{2\pi} d\tilde{\phi}
\int_{0}^{2/\tilde{\gamma}_{b}} d\tilde{\theta}
\int_{\tilde{X}_{\rm inner}(\tilde{\theta}, \tilde{\gamma}_{b})}
^{\tilde{X}_{\rm outer}(\tilde{\theta}, \tilde{\gamma}_{b})} 
y^{2}\xi(\mid \vec{r}-\tilde{\vec{r}}\mid)
g(\tilde{x}, \tilde{\theta}, \tilde{\gamma}_{b})d\tilde{x} ,
\label{eq:integ2}
\end{eqnarray}
where 
\begin{eqnarray}
\mid \vec{r}-\tilde{\vec{r}}\mid & =& [(xR_{b}\sin\theta-
\tilde{x}\tilde{R}_{b}\sin\tilde{\theta}\cos\tilde{\phi})^{2}
+(\tilde{x}\tilde{R}_{b}\sin\tilde{\theta}\sin\tilde{\phi})^{2} \nonumber \\
 & & +(xR_{b}\cos\theta-\tilde{x}\tilde{R}_{b}\cos\tilde{\theta})^{2}]^{1/2} .
\label{eq:dis12}
\end{eqnarray}

Using the above integrals we may write
\begin{equation}
\zeta(\tau)=\langle \Delta(T) \Delta(T+\tau) \rangle=I_{2}/I_{1} .
\label{eq:I2}
\end{equation}

\section{NUMERICAL RESULTS}

We evaluated numerically the integrals in equations~(\ref{eq:integ1}) and
(\ref{eq:integ2}) using a Gaussian quadrature method. For each of the two 
different frequency regions ($\nu<\nu_{\rm peak}$ 
and $\nu>\nu_{\rm peak}$), we considered six different
cases with values of $\gamma_{b}$ of 100, 22 and 3, which correspond to
peak afterglow emission in the $X$-ray, optical and radio wavelength
regimes, and for each $\gamma_{b}$, we considered two different cases of
$r_{0}/R_{b}=10^{-3}$ and $r_{0}/R_{b}=10^{-2}$. The values of
$\gamma_{b}$, $R_{b}$ and $\nu_{\rm peak}$, the frequency at which the
observed flux peaks at a given observation time $T_{{\rm obs}}$, 
are given by (Granot et al. 1999),
\begin{equation}
\gamma_{b} = 7.96\left( \frac{E_{52}}{n_1}\right)^{1/8}
\left( \frac{T_{{\rm obs,days}}}{1+z} \right)^{-3/8} ,
\label{eq:gamma_b}
\end{equation}
\begin{equation}
R_{b} = 3.29\times 10^{17} \left[ \frac{E_{52}T_{{\rm obs,days}}}
{n_1(1+z)} \right]^{1/4} \mbox{cm} ,
\label{eq:R_b}
\end{equation}
\begin{equation}
\nu_{\rm peak} = 7.29\times 10^{15}\sqrt{1+z} \frac{\phi_{{\rm
peak}}(p)}{\phi_{{\rm peak}}(2.5)}\frac{f(p)}{f(2.5)}
\epsilon^{1/2}_{B}\epsilon^{2}_{e}E^{1/2}_{52}T^{-3/2}_{{\rm
obs,days}}~\mbox{Hz} .
\label{eq:fre_peak}
\end{equation}
where $\phi_{{\rm peak}}$ is a slowly decreasing function of $p$,
$f(p)\equiv[(p-2)/(p-1)]^2$, and $T_{{\rm obs,days}}$ is the observation 
time in days and is related to $T$ by equation~(\ref{eq:obs_time}).

In our numerical calculations we assume $p=2.5$ (Granot et al. 1999, 
Sari, Piran \& Halpern 1999), $\epsilon_{B}=0.1$,
$\epsilon_{e}=0.1$ and $\eta=0.1$.  Table~1 shows the associated $T$,
$T_{\rm obs}$, $R_{b}$ and $\nu_{\rm peak}$ for the three choices of
$\gamma_{b}$, with $E_{52}$, $n_1$ and $z$ as free parameters.

Figures~3 and 4 show the numerical results for the six cases mentioned
above, with Figure~3 corresponding to $\nu<\nu_{\rm peak}$ and Figure~4
corresponding to $\nu>\nu_{\rm peak}$.  We plot the square root of the
autocorrelation function for the temporal fluctuations of the afterglow
flux $\zeta(\tau_{\rm obs})$, as normalized by the unknown amplitude of the
fractional density fluctuations in the ambient medium, $\xi_{0}^{1/2}$. We
normalize $r_{0}$ by $R_{b}$, which is different for the three different
cases of $\gamma_{b}$.

The value of $[\zeta(\tau_{\rm obs})]^{1/2}$ at $\tau_{\rm obs}=0$ provides
the typical amplitude of the fluctuations in the observed flux.  The
charateristic period of the fluctuations is $\tau_{1/2,{\rm obs}}$, the
time over which $\zeta(\tau_{\rm obs})$ drops to half its maximum value.
We list the derived values of these parameters in Tables~2 and 3.  The
tables implies that: (i) for a given $r_{0}/R_{b}$,
$[\zeta(0)/\xi_{0}]^{1/2}$ decreases with decreasing $\gamma_{b}$; and (ii)
for a given $\gamma_{b}$, $[\zeta(0)/\xi_{0}]^{1/2}$ decreases with
decreasing $r_{0}/R_{b}$.  Poisson statistics implies that density
fluctuations with an amplitude $\delta_0\ll 1$ on a correlation length $l$
will induce an average fluctuation amplitude $\sim \delta_0 / \sqrt N$ in a
region of size $L\gg l$, where $N \sim (L/l)^3$ is the number of
independent regions of positive or negative density fluctuations in the
sampled volume.  This explains the qualitative trend of $[\zeta(0)]^{1/2}$
to decrease as the value of $r_0$ is lowered. However, $[\zeta(0)]^{1/2}$
is not proportional to $r_{0}^{3/2}$ in our problem, because the different
uncorrelated regions within the integration volume have different weights
in their contribution to the total flux.

\section{CONSTRAINTS FROM GRB990510}

Stanek et al. (1999) have monitored the optical afterglow of GRB990510 on a
sub-hour basis and obtained an {\it rms} scatter of 0.02 mag for the $BVRI$
observations during its first day, consistent with the scatter being
entirely due to photometric noise (see also Hjorth et al.  1999). These
observations imply an upper limit of $2\%$ on the {\it rms} amplitude of
optical flux fluctuations on time scales from 0.1 to 2 hours during the
first day of the afterglow.  After 1.6 days, the afterglow decline rate
steepened, possibly due to the lateral expansion of a jet (Stanek et
al. 1999; Harrison et al. 1999).  Lateral expansion is expected to be
important when the Lorentz factor decelerates to a value of order the
inverse of the jet opening angle, but could be neglected at earlier times
(Rhoads 1997, 1999). In the following, we focus on the early stage of this
afterglow ($<1.6$ days), during which the observed afterglow radiation
originates from a region much smaller than the jet opening angle. At this
stage, the observed region behaves as if it is part of a
spherically-symmetric fireball, and hence should be adequately decribed by
our spherical expansion model.  Our model also assumes that the unperturbed
ambient gas has a uniform density.  Chevalier \& Li (1999) studied wind
interaction models for GRB afterglows, and concluded that GRB990510 can be
better explained by a constant density medium than a wind density profile.

The equivalent $\gamma$-ray energy release of GRB990510 for isotropic
emission is $E=1.2\times10^{53}$erg, if the source redshift is $z=1.62$
(Wijers et al. 1999).  We assume that the hydrodynamic energy release is
comparable to this value, and also adopt $n=1~{\rm cm^{-3}}$.  The
afterglow emission peaks in the optical ($\nu_{{\rm
peak}}=5.1\times10^{14}$Hz) at $T_{\rm obs}=9.6$hr, when $\gamma_b= 22$ and
$R_{b}=3.9\times10^{17}$cm.  Since most of the observational data from
Stanek et al. (1999) is at frequencies $\nu>\nu_{\rm peak}$, we compute the
flux fluctuations in this regime.  Our calculations indicate that flux
fluctuations on time scales ($\tau_{1/2,{\rm obs}}$) between 0.3 and 2
hours correspond to density fluctuations on length scales, $r_{0}$, between
$3.5\times 10^{14}$ and $2.6\times 10^{15}$cm, i.e. in the range of
$\sim$20--200 AU.  Based on the observed upper limit on the amplitude of
flux fluctuations in GRB990510, we calculate the upper limit on
$\xi_{0}^{1/2}$ as a function of $r_{0}$ in this range.  The resulting
constraints on the ISM inhomogenties in the vicinity of the progenitor of
GRB990510 are illustrated by the solid line in Figure~5.  The
horizontally-shaded region above this curve is forbidden while the region
below the curve is allowed, based on the afterglow data.  We find that the
amplitude of density fluctutations has to be lower than $\sim 10\%$ on the
length scale of $\sim 200$ AU and lower than unity on the $\sim 20$ AU
scale.

It is instructive to compare our derived constraints with observational
data for density fluctuations on similar length scales in the ISM of the
Milky-Way galaxy.  Structure on small scales was first inferred by Dieter,
Welch, \& Romney (1976), using VLBI observations at 21 cm against the
extragalctic source 3C 147.  Diamond et al. (1989) and Davis, Diamond, \&
Goss (1996) obtained similar results for more sources.  Frail et al.
(1994) detected temporal variations in the 21 cm absorption towards six
high-velocity pulsars, and inferred changes in the HI column-density of
$\sim 13\%$ on projected scales of 5--100 AU.  This sets a lower limit on
the {\it rms} density contrast $(\delta n_H/n_H)$ of $\sim 13\%$ in spheres
of the above scale; the actual density contrast could be much larger due to
partial cancellations between overdense and underdense regions along the
line-of-sight to the pulsars.  Lauroesch \& Meyer (1999) studied the small
scale ISM structure in atomic gas by observing the interstellar K {\small
I} absorption line towards multiple star systems, and inferred a hydrogen
density contrast $(\delta n_H/n_H) \sim 1-2$ on the length scales of
$10^2$--$10^3$AU (Meyer 1999, private communication).  Small-scale density
inhomogenieties were also inferred in molecular clouds (Marscher, Moore, \&
Bania, 1993; Moore \& Marscher, 1995).  All these studies indicate that
structure is ubiquitous on scales of 10--$10^3$ AU in the ISM (for a
physical interpretation of the above results, see Heiles 1997). The
observed structure might be caused in part by fluctuations in the
ionization fraction or chemistry. However, for the purpose of putting our
results in the context of these local ISM observations, we will assume that
they relate to actual inhomogneities in the gas density.  In order to
compare our results with the above data we need a relation between the
autocorrelation function and the {\it rms} amplitude of density contrast in
a region of a given size.  For a spherical region of radius $R$, this
relation is to a good approximation given by (Padmanabhan 1993)
\begin{equation}
\left \langle \left( \frac{\delta n}{n} \right)^2_R \right \rangle
\cong \frac{1}{2\pi^2 V} \int_{0}^{R^{-1}} k^3 \sigma_{k}^2
\frac{dk}{k} ,
\label{eq:padmanabhan}
\end{equation}
where $V$ is the normalization volume, $\sigma_{k}^2$ is the power spectrum
of the density fluctuation and is equal to the Fourier transfer of the
autocorrelation function:
\begin{equation}
\sigma_{k}^2 = V \int \xi(\vec{r})e^{-i\vec{k}\cdot \vec{r}}d^3\vec{r} .
\label{eq:powerspect}
\end{equation}
For the autocorrelation function in equation~(\ref{eq:auto}), 
\begin{eqnarray}
\sigma_{k}^2 & = & V \int_{0}^{\infty}dr \int_{0}^{\pi}d\theta
\int_{0}^{2\pi}d\phi \xi_0 e^{-r/r_0} e^{-ikr\cos\theta}
r^2\sin\theta \\
 & = & 4\pi\xi_{0} V \int_{0}^{\infty}\frac{r}{k}e^{-r/r_0}\sin krdr .
\label{eq:powerspect1}
\end{eqnarray}
Subtituting this result into equation~(\ref{eq:padmanabhan}), we get
\begin{equation}
\left \langle \left( \frac{\delta n}{n} \right)^2_R \right \rangle 
 =  \frac{2}{\pi}\xi_0 \int_{0}^{\infty}re^{-r/r_0}dr
\int_{0}^{R^{-1}}k\sin krdk 
 =  \frac{2}{\pi}\xi_0 \left[\arctan \left( \frac{r_0}{R} \right)
-\frac{R/r_0}{1+(R/r_0)^2} \right] .
\label{eq:contrast}
\end{equation}
Based on observational constraints for $\langle (\delta n / n )^2_R
\rangle$ on a given length scale $R$, we may now calculate from
equation~(\ref{eq:contrast}) the related value of $\xi_0^{1/2}$ for any
assumed value of $r_0$.

The short-dashed curve in Figure~5 shows the constraint imposed by setting
$\langle (\delta n / n )^2_R \rangle = 1$ for $R=10^2$AU, while the
long-dashed curve corresponds to $\langle (\delta n / n )^2_R \rangle = 1$
for $R=10^3$AU. The vertically-shaded region between these two curves
describes intermediate length scales, refering to the range of constraints
imposed by the observations of Lauroesch \& Meyer (1999). This entire
region is well above the solid curve calculated from GRB990510, implying
that we should not have been surprised if flux fluctuations at a level of a
few percent were detected in the optical afterglow of GRB990510.  The
dot-dashed curve in this plot corresponds to setting $\langle (\delta n / n
)^2_R \rangle^{1/2} = 0.13$ for $R=10^2$AU, which refers to the lower limit
on the HI density fluctuations inferred by Frail et al. (1994). The allowed
range of density fluctuations are above this curve. This constraint is
close but still above the solid curve, i.e. in the region excluded by the
variability data on GRB990510.  In summary, we find that if a GRB occurs in
the ISM where the density perturbations have similar properties to those
inferred by Lauroesch \& Meyer (1999) or Frail et al. (1994), then the
resulting fluctuations in the afterglow flux should be detectable.

\section{CONCLUSIONS}

We have found that linear density fluctuations with $\delta n/n\la 1$ on
the length scale of $\sim 1$--$10^3$ AU could induce afterglow flux
fluctuations with a fractional amplitude of up to $\sim40\%$ over
timescales of tens of seconds in the X-rays, up to $\sim30\%$ over tens of
minutes in the optical, and up to $\sim9\%$ over days in the radio (see
Table 3). These flux fluctuations average over the full range of density
inhomoegeneities within the emission region.  For example, during the
optical afterglow, the emission region occupies a rather large volume of
$\sim (10^4~{\rm AU})\times (10^3~{\rm AU}) \times (10^3~{\rm AU})$
(assuming $E_{52}=1$, $n_1=1$, $z=1$) and so inhomogeneities on scales~$\ll
10^3~{\rm AU}$~would surely be ensemble averaged as long as their volume
filling fraction is not too small.  

At both extremes of high (X-ray) and low (radio) frequencies, the
calculated variability might be contaminated by other effects. During the
early period of the X-ray afterglow, the externally-induced fluctuations we
considered might be blended with variability associated with internal
shells within the fireball which are still catching-up with the
decelerating blast wave, and which were ignored in our analysis.  At the
opposite extreme of low radio frequencies, the flux might scintillate due
to inhomogeneities in the local (Galactic) interstellar medium along the
line-of-sight (Goodman 1998; Frail et al. 1997, 1999; Waxman et al. 1998).
In principle, the radio flux variability predicted by our model can be
distinguished from variability due to scintillations, based on its
different dependence on photon frequency, especially at high frequencies.
However, it appears that the best spectral regime to observe the afterglow
flux variability predicted in this paper is in between the X-ray and radio
frequency windows, e.g. in the optical-infrared band.

Our calculation assumed spherical symmetry and should apply to the early
expansion stages of a jet, as long as its Lorentz factor is larger than the
inverse of its opening angle (Rhoads 1997, 1999, Sari et al. 1999). 
Our treatment could be extended in the future to describe the
lateral expansion of a jet at later times, the possible existence of a
power-law density profile as for a precurser wind from the GRB progenitor
(Chevalier \& Li 1999), and the possible effects of nonlinear clumps of
density (Dermer \& Mitman 1999).

The application of our simple model to the early optical afterglow of
GRB990510 (Stanek et al. 1999) provides already interesting upper limits on
the density fluctuations on scales of $\sim 20$--$200$ AU around the
source.  These limits are {\it lower} than the observed fluctuation
amplitude on similar scales in the local interstellar medium (Fig. 5).  If
these local measurements apply to interstellar turbulence in high redshift
galaxies, then optical monitoring of future afterglows should reveal flux
fluctuations at the level of a few percent or higher on the timescale of
less than an hour.

\acknowledgments

We thank Bruce Draine, Dale Frail and David Meyer for useful discussions
about the detection of density fluctuations in the ISM.  This work was
supported in part by NASA grants NAG 5-7039 and NAG 5-7768.

\vfil\eject
\begin{table}
\begin{center}
\begin{tabular}{|c|c|c|c|c|}                              \hline
$\gamma_{b}$ & $T_{\rm obs}/[E_{52}^{1/3}n_{1}^{-1/3}((1+z)/2)]$ &
$R_{b}/(E_{52}^{1/3}n_{1}^{-1/3})$ & 
$\nu_{\rm peak}/[n_{1}^{1/2}((1+z)/2)^{-1}]$ \\ \hline
100 & 202.6 sec & $6.1\times10^{16}$~cm & $2.9\times10^{17}$~Hz \\ \hline 
22 & 3.2 hr & $1.7\times10^{17}$~cm & $6.7\times10^{14}$~Hz \\ \hline 
3 & 27.0 day & $6.3\times10^{17}$~cm & $2.3\times10^{11}$~Hz \\ \hline
\end{tabular}
\end{center}
\begin{center}
\caption{Values of $T_{\rm obs}$, $R_{b}$ and $\nu_{\rm peak}$ for the three
choices of $\gamma_{b}$. $E_{52}$, $n_1$ and $z$ are left as free parameters.}
\end{center}
\label{table:1}
\end{table}

\begin{table}
\begin{center}
\begin{tabular}{|c|c|c|c|c|c|c|}                              \hline
case & $\gamma_{b}$ & $r_{0}/R_{b}$ & $r_{0}/(E_{52}^{1/3}n_{1}^{-1/3})$ &
$[\zeta(0)/\xi_{0}]^{1/2}$ & $\tau_{1/2,{\rm obs}}/T_{\rm obs}$ & 
$\tau_{1/2,{\rm obs}}/[E_{52}^{1/3}n_{1}^{-1/3}((1+z)/2)]$ \\ \hline 
1 & 100 & 0.001 & $6.1\times10^{13}$~cm & 0.05 & 0.026 & 5.3~sec \\ \hline
2 & 100 & 0.01 & $6.1\times10^{14}$~cm & 0.28 & 0.30 & 61.6~sec \\ \hline 
3 & 22 & 0.001 & $1.7\times10^{14}$~cm & 0.017 & 0.027 & 0.086~hr \\ \hline
4 & 22 & 0.01 & $1.7\times10^{15}$~cm & 0.22 & 0.30 & 0.94~hr \\ \hline 
5 & 3 & 0.001 & $6.3\times10^{14}$~cm & 0.004 & 0.026 & 0.70~day \\ \hline
6 & 3 & 0.01 & $6.3\times10^{15}$~cm & 0.07 & 0.26 & 7.0~day \\ \hline 
\end{tabular}
\end{center}
\begin{center}
\caption{Values of $[\zeta(0)/\xi_{0}]^{1/2}$,
$\tau_{1/2,{\rm obs}}/T_{\rm obs}$ and $\tau_{1/2,{\rm obs}}$ for six cases 
in the $\nu<\nu_{\rm peak}$ region. 
The values of $E_{52}$, $n_1$ and $z$ are left as free parameters.}
\end{center}
\label{table:2}
\end{table}

\begin{table}
\begin{center}
\begin{tabular}{|c|c|c|c|c|c|}                              \hline
case & $\gamma_{b}$ & $r_{0}/R_{b}$ &
$[\zeta(0)/\xi_{0}]^{1/2}$ & $\tau_{1/2,{\rm obs}}/T_{\rm obs}$ & 
$\tau_{1/2,{\rm obs}}/[E_{52}^{1/3}n_{1}^{-1/3}((1+z)/2)]$ \\ \hline 
1 & 100 & 0.001 & 0.06 & 0.033 & 6.7~sec \\ \hline
2 & 100 & 0.01 & 0.37 & 0.33 & 67~sec \\ \hline 
3 & 22 & 0.001 & 0.026 & 0.034 & 0.11~hr \\ \hline
4 & 22 & 0.01 & 0.27 & 0.31 & 1.0~hr \\ \hline 
5 & 3 & 0.001 & 0.009 & 0.031 & 0.84~day \\ \hline
6 & 3 & 0.01 & 0.09 & 0.24 & 6.5~day \\ \hline 
\end{tabular}
\end{center}
\begin{center}
\caption{Values of $[\zeta(0)/\xi_{0}]^{1/2}$,
$\tau_{1/2,{\rm obs}}/T_{\rm obs}$ and $\tau_{1/2,{\rm obs}}$ for six cases 
in the $\nu>\nu_{\rm peak}$ region. 
The values of $E_{52}$, $n_1$ and $z$ are left as free parameters.}
\end{center}
\label{table:3}
\end{table}

\clearpage

\begin{figure}
\includegraphics{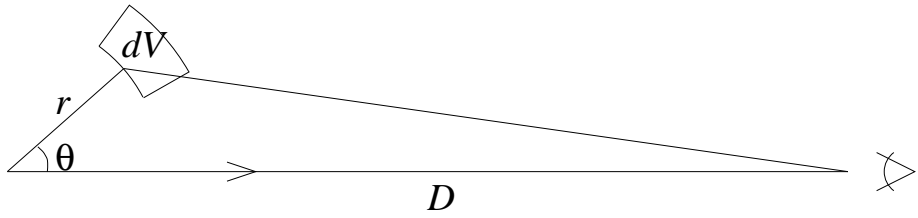}
\vspace{2.1in}
\caption{Coordinate system for the integration of the emitted afterglow
flux.}
\label{fig:1}
\end{figure}

\vfil\eject

\begin{figure}
\includegraphics{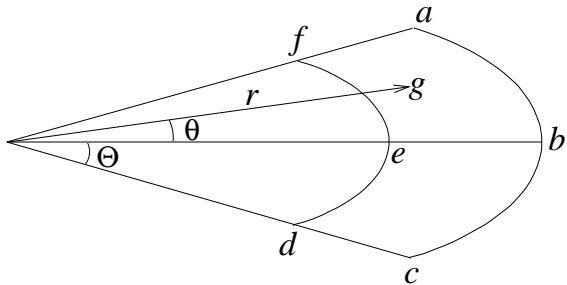}
\vspace{2.1in}
\caption{Notations and geometry of the integration region.}
\label{fig:2}
\end{figure}

\vfil\eject

\begin{figure}[t]
\includegraphics{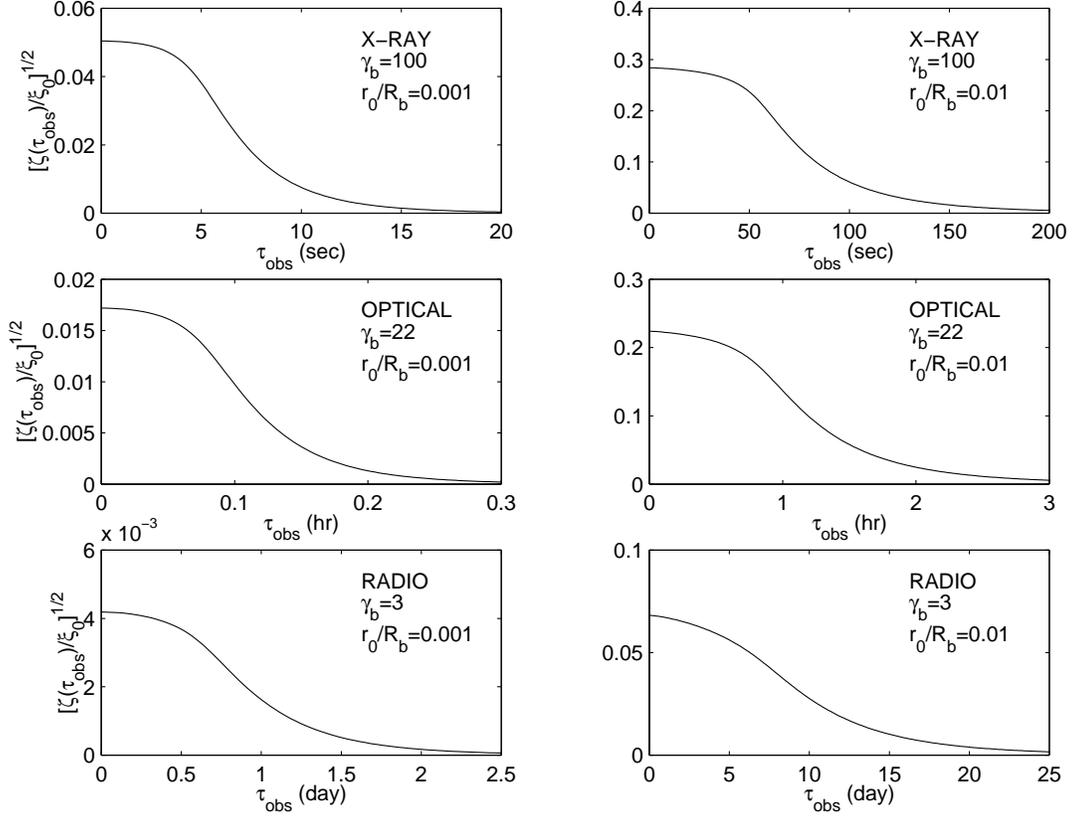}
\vspace{4.5in}
\caption{The temporal correlation function for variations of the aftreglow
flux $[\zeta(\tau_{\rm obs})/\xi_{0}]^{1/2}$ as a function of $\tau_{\rm
obs}$ for $\nu<\nu_{\rm peak}$, where $\tau_{\rm obs}$ scales as
$[E_{52}^{1/3}n_{1}^{-1/3}((1+z)/2)]$.  The Six plots correspond to six
cases for different values of parameters, as shown next to each plot.}
\label{fig:3}
\end{figure}

\vfil\eject

\begin{figure}[t]
\includegraphics{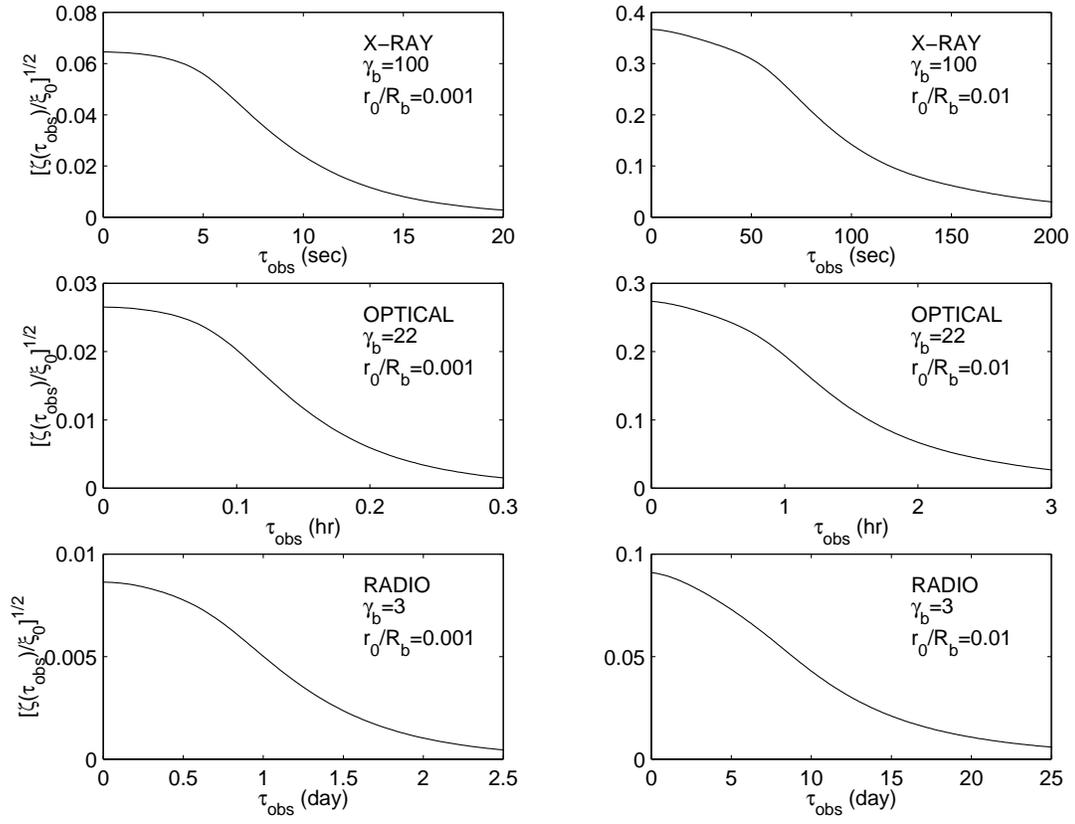}
\vspace{4.5in}
\caption{Same as in Figure 3, but for $\nu>\nu_{\rm peak}$.}
\label{fig:4}
\end{figure}

\vfil\eject

\begin{figure}[t]
\includegraphics{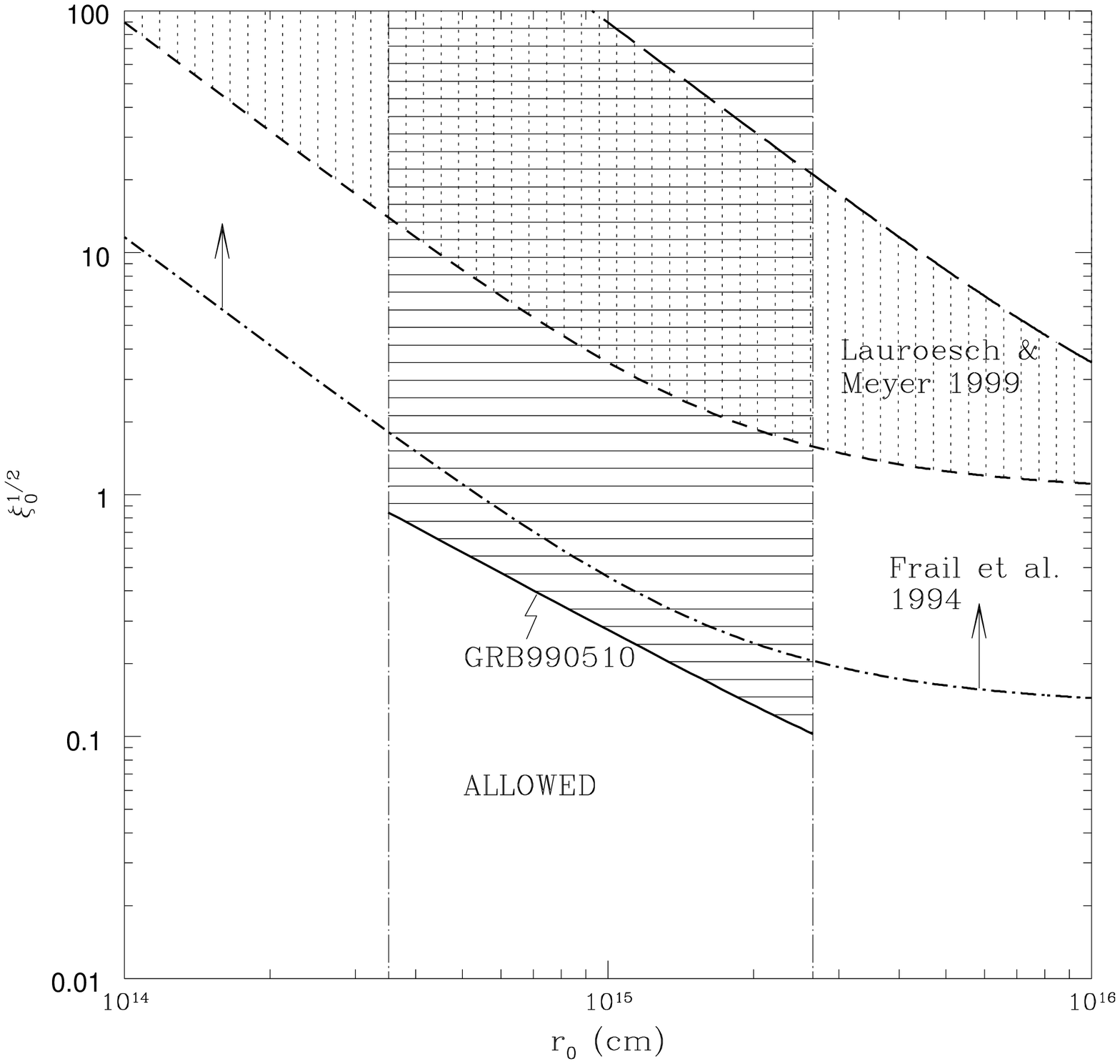}
\vspace{4.5in}
\caption{Constraints on the $\xi_{0}$--$r_{0}$ parameter space, for a
density autocorrelation function of the form $\xi(r)=\xi_0\exp(-r/r_0)$.
The solid curve describes the upper limit on $\xi_{0}^{1/2}$ for $r_{0}$
between $4\times 10^{14}$ to $3\times 10^{15}$cm based on the upper limit
on the amplitude of optical flux fluctuations in the GRB990510 afterglow
(Stanek et al. 1999). The upper horizontaly-shaded region is not allowed by
the GRB990510 data.  The short-dashed curve and the long-dashed curve
reflect constraints on the inhomogeneities in the local ISM, based on
Lauroesch \& Meyer (1999), with the short-dashed curve corresponding to
$\langle (\delta n / n )^2_R \rangle = 1$ in spheres of radius $R=10^2$AU,
and the long-dashed curve corresponding to $\langle (\delta n / n )^2_R
\rangle = 1$ in spheres of radius $R=10^3$AU. The vertically-shaded region
between these two curves describes intermediate length scales.  The
dot-dashed curve corresponds to $\langle (\delta n / n )^2_R \rangle^{1/2}
= 0.13$ in spheres of radius $R=10^2$AU and describes the lower limit on
$\xi_{0}^{1/2}$ inferred from Frail et al. (1994).}
\label{fig:5}
\end{figure}


\begin{references}

\reference{} Blandford, R. D., \& McKee, C. F. 1976, Phys. Fluids, 19, 1130

\reference{} Chevalier, R. A., \& Li, Z. Y. 1999, ApJ, submitted
(astro-ph/9908272)

\reference{} Davis, R. J., Diamond, P. J., \& Goss, W. M. 1996, MNRAS, 283,
1105

\reference{} Dermer, C., \& Mitman, K. 1999, ApJ, 513, L5

\reference{} Diamond, P. J. Goss, W. M., Romney, J. D., Booth, R. S., 
Kalberla, P. N. W., \& Mebold, U. 1989, ApJ, 347, 302

\reference{} Dieter, N. H., Welch, W. J., \& Romney, J. D. 1976, ApJ, 206,
L113

\reference{} Frail, D. A., Kulkarni, S. R., Nicastro, L., Feroci, M., \& 
Taylor, G. B. 1997, Nature, 389, 261

\reference{} Frail, D. A., Waxman, E., \& Kulkarni, S. R. 1999, ApJ, submitted 
(astro-ph/9910319)

\reference{} Frail, D. A., Weisberg, J. M., Cordes, J. M., \& Mathers,
C. 1994, ApJ, 436, 144

\reference{} Goodman, J. 1997, New Astronomy, 2, 449

\reference{} Granot, J., Piran, T., \& Sari, R. 1999, ApJ, 513, 679

\reference{} Harrison, F. A., et al. 1999, ApJ, 523, L121

\reference{} Heiles, C. 1997, ApJ, 481, 193

\reference{} Hjorth, J., et al. 1999, GCN Circ. 320

\reference{} Lauroesch, J. T., \& Meyer, D. M. 1999, ApJ, 519, L181

\reference{} Marscher, A. P., Moore, E. M., \& Bania, T. M. 1993, 
ApJ, 419, L101

\reference{} M\'{e}sz\'{a}ros, P., \& Rees, M. J. 1993, ApJ, 405, 278

\reference{} M\'{e}sz\'{a}ros, P., \& Rees, M. J. 1997, ApJ, 476, 232

\reference{} Moore, E. M., \& Marscher, A. P. 1995, ApJ, 452, 671

\reference{} Paczy\'{n}ski, B., \& Rhoads, J. E. 1993, ApJ, 418, L5

\reference{} Padmanabhan, T. 1993, Structure Formation in the Universe
(Cambridge: Cambridge University Press), p. 200

\reference{} Piran, T. 1999, Phys. Reports 314, 575

\reference{} Reif, F. 1965, Fundamentals of Statistical and Thermal Physics 
(New York: McGraw-Hill), pp. 583--585

\reference{} Rhoads, J. E. 1997, ApJ, 487, L1

\reference{} Rhoads, J. E. 1999, ApJ, in press (astro-ph/9903399)

\reference{} Rybicki, G. B., \& Lightman, A. P. 1997, Radiative Processes 
in Astrophysics (New York: Wiley Interscience), p. 147

\reference{} Sari, R., Piran, T., \& Halpern, J. P. 1999, ApJ, 519, L17

\reference{} Sari, R., Piran, T., \& Narayan, R. 1998, ApJ, 497, L17

\reference{} Stanek, K. Z., Garnavich, P. M., Kaluzny, J., Pych, W., Thompson, I. 
1999, ApJ, 522, L39

\reference{} Waxman, E. 1997a, ApJ, 485, L5

\reference{} Waxman, E. 1997b, ApJ, 489, L33

\reference{} Waxman, E. 1997c, ApJ, 491, L19

\reference{} Waxman, E., Kulkarni, S. R., \& Frail, D. A. 1998, ApJ, 497, 288

\reference{} Wijers, R. A. M. J., Rees, M. J., \& M\'{e}sz\'{a}ros, P. 
1997, MNRAS, 288, L51

\reference{} Wijers, R. A. M. J., \& Galama, T. J. 1998, ApJ, submitted
(astro-ph/9805341)

\reference{} Wijers, R. A. M. J., et al. 1999, ApJ, submitted (astro-ph/9906346)

\end{references}
\end{document}